\documentstyle[preprint,aps]{revtex}
\begin{document}
\draft
\title{The Dendritic magnetic avalanches in carbon-free MgB$_2$ thin films with and without a deposited Au layer}
\author{Eun-Mi Choi,$^1$ Hyun-Sook Lee,$^1$ Hyun Jung Kim,$^1$ Byeongwon Kang,$^1$ and Sung-Ik Lee$^{1,2}$}
\address{$^1$National Creative Research Initiative Center for Superconductivity, Department of Physics,
Pohang University of Science and Technology, Pohang 790-784,
Republic of Korea}
\address{$^2$Quantum Material{'}s Research Laboratory, Korea Basic Science Institute, Daejeon 305-333, Korea}
\author{A. A. F. Olsen,$^3$ D. V. Shantsev$^3$ and T. H. Johansen$^{3,4}$}
\address{$^3$Department of Physics, University of Oslo, PO Box 1048 Blindern, N-0316 Oslo, Norway }
\address{$^4$Texas Center for Superconductivity and Advanced Materials, University of Houston, Houston, Texas 77204-5002, USA}
\maketitle

\begin{abstract}

From the magneto optics images (MOI), the dendritic magnetic
avalanche is known to appear dominantly for thin films of the newly
discovered MgB$_2$. To clarify the origin of this phenomenon, we
studied in detail the MOI of carbon-free MgB$_2$ thin films with and
without a deposited gold layer. The MOI indicated carbon
contamination was not the main source of the avalanche. The MOI
clearly showed that the deposition of metallic gold deposition on
top of a MgB$_2$ thin film improved its thermal stability and
suppressed the sudden appearance of the dendritic flux avalanche.
This is consistent with the previous observation of flux noise in
the magnetization.
\end{abstract}

\pacs{} The binary metallic MgB$_2$ superconductor with $T_c$ = 39 K
is a very interesting material for basic science and applications.
The relatively high upper critical field $H_{c2}$(20 - 30 T, 49
T)\cite{Jung,Gurevich} of MgB$_2$ and its extremely high critical
current density $J_c$($\sim$ 10$^7$ A/cm$^2$)\cite{Hyeong},
especially in thin films, suggest that MgB$_2$ could be a much more
important superconducting material compared to conventional metallic
superconductors. However, the critical current was found to be
seriously limited by a phenomenon called the vortex avalanche (flux
jumps or magnetic flux noise). In particular, this phenomenon
prevails at low temperatures ($T$ $<$ 15 K) and low magnetic fields
($H\leq$ 1000 Oe)\cite{Johansen2,Wen}. The dendritic flux patterns
formed in the course of an avalanche were studied in detail by using
magneto optic imaging (MOI)\cite{Johansen2}. Even though the details
were somewhat different, abrupt flux dynamics with dendritic
penetration had been observed earlier in other superconductors such
as Nb films\cite{Nb} and YBCO films, where a laser pulse is needed
to trigger the instability\cite{Leiderer}.

The dendritic flux penetration can be explained by a thermomagnetic
instability due to the heat generated by vortex motion. Aranson $et$
$al.$\cite{Aranson2} and Rakhmanov $et$ $al.$\cite{Rakhmanov}
carried out a linear analysis and found a numerical solution to the
thermal diffusion and Maxwell equations, predicting that an
instability might result in very nonuniform dendritic-like
temperature and flux distributions. They also found a criterion for
the onset of the instability as well as its build-up time.

Experimentally, Choi $et$ $al.$ reported that the flux jumps in
MgB$_2$ films could be cured by the superconductors being thermally
stabilized after having been coated with a metallic Au
film\cite{choi}. They observed, for a gold coating, a dramatic
reduction in the flux noise in the magnetic hysteresis ($M-H$) loop
and an enhancement of $J_c$ up to 1.22 $\times$ 10$^7$ A/cm$^2$ at
$H$ = 0 Oe and $T =$ 5 K\cite{choi}. This strongly suggested that
thermal stabilization greatly suppressed the vortex avalanche.

Recently Ye $et$ $al.$\cite{Ye} reported MOI results and $M-H$
curves for two kinds of MgB$_2$ thin film samples: one without
serious carbon contamination and one with 12\% carbon contamination
(C$_{0.12}$MgB$_2$, called C-MgB$_2$ in this paper). Both were
prepared by using the hybrid chemical-physical vapor deposition
(HPCVD). Since dendrites were seen only in C-MgB$_2$, they concluded
that the avalanches mainly involved electron scattering due to
carbon contamination, and were not the result of a thermal
instability.

To clarify the origin of the avalanche behavior, we synthesized
carbon-free $\it c$-axis oriented MgB$_2$ thin films with and
without a gold coating and performed MOI, and magnetization
measurements. In this work, we showed that ultra clean, carbon-free
MgB$_2$ thin films, indeed, displayed dendritic flux avalanches;
moreover, we found that the avalanches disappeared with the addition
of a gold coating, all of which support the thermal instability
scenario and not electron scattering due to carbon doping.

The MgB$_2$ thin films in this study were fabricated by using a two
step method\cite{Kang}. Briefly, an amorphous boron thin film was
deposited on a Al$_2$O$_3$ ${(1 \ \bar{1} \ 0 \ 2)}$ substrate at
room temperature by using pulsed laser deposition. The base pressure
was 10$^{-8}$ Torr. The B film was put into a Nb tube with
high-purity Mg (99.9$\%$) and heat treated. To eliminate possible
contamination with oxygen, water, and carbon, we never allowed the
samples to be exposed to air until the final thin film had been
produced. Post-annealing was done in a high-purity Ar (99.999$\%$)
atmosphere.

The carbon contents in these samples were confirmed to be below the
resolution limits of wavelength dispersive spectroscopy (WDS,
0.01\%) and energy dispersive spectroscopy (EDS, 1\%) by using
electron probe micro analysis (EPMA). These values were far below
the 12\% for C-MgB$_2$ prepared by Ye $et$ $al.$

Rectangular samples with dimensions of 2 $\times$ 3 mm$^2$ and with
thickness of 500 nm were chosen for the magnetic measurements. The
$\it c$-axis orientation of the MgB$_2$ film was confirmed using
scanning electron microscopy (SEM)\cite{Kang3}. The magnetization
($M-T$ and $M-H$ curves) was measured by using a SQUID magnetometer
(Quantum Design, MPMSXL) for $H \| c$ axis. In this experiment, we
measured the $M-H$ loop (at $T =$ 5 K) and the MOI (at $T =$ 3.8 K)
of MgB$_2$ thin films coated with Au films of different thicknesses.
The gold deposition caused no observable deterioration in the
superconductivity of the samples\cite{choi}. MOI was performed using
a film of in-plane magnetization ferrite garnet as the
Faraday-active sensor measuring the field distribution over the
surface of the MgB$_2$ sample. The sample was mounted on the cold
finger in an Oxford Microstat-He optical cryostat placed under a
polarized light microscope. The details of the MOI setup can be
found elsewhere\cite{Johansen}.

Figure 1 shows the temperature dependent low-field magnetic
susceptibility for a bare MgB$_2$ thin film, which has a sharp
transition with onset at $T_c =$ 39 K. The superconducting
transition for the MgB$_2$ film coated with a thick Au layer was the
same as that shown in Fig. 1, implying that the gold deposition did
not adversely affect the superconducting properties. For the
elemental analysis, we measured wavelength dispersive spectra for
many different surfaces by using EPMA (inset in Fig. 1). The inset
shows that, within the resolution limit, carbon was not present in
the MgB$_2$ films. We could only detect B, Mg, O, and Al, where the
signals from O and Al are from the Al$_2$O$_3$ substrate.

Figure 2 (a) shows MOI of the flux penetration in a carbon-free
MgB$_2$ thin film with the magnetic field perpendicular to the film
at $H =$ 34 mT and $T =$ 3.8 K. One can see that the flux penetrates
the film in a typical dendritic fashion. It should be emphasized
that the exact dendrite pattern varies widely between experiments
repeated under the same external conditions; hence, the structure is
not a fingerprint of defect regions in the superconductor. Also, the
extremely abrupt formation of each flux dendrite, faster than we
could capture with our image recording system, demonstrates a
characteristic feature of an instability, presumably one of
thermo-magnetic origin.

For films made by using HPCVD, the dendrites appear only in
C-contaminated MgB$_2$ thin films, which is quite different from our
observation. The main conclusion of Ye $et$ $al.$ for films prepared
using HPCVD was that magnetic flux jumping was mostly due to
electron scattering by carbon impurity, but that cannot be correct
because flux jumping still was presented in our carbon-free MgB$_2$
thin films.

To clarify this issue, we prepared MgB$_2$ films with Au coatings of
three different thicknesses: 0.2, 0.9, and 2.55 $\mu m$. The MO
images obtained at 3.8 K for these three films are shown in Fig. 2
(b-d), respectively. It is evident that with increasing Au
thickness, the dendritic character of the flux penetration gradually
disappears. We notice, however, that even with the thickest Au
layer, the flux penetration pattern is far from being ideally
smooth. The reason for this is the presence of small defects being
spread over the films area and along the edge, and distorting the
penetration by creating well-known parabolic or fan-like flux
patterns. These patterns are very easily distinguished from the
dendritic structures by their vastly different dynamics. All the
irregular flux patterns seen in Fig. 1 (d) evolve slowly when the
applied field is ramped at a slow rate. The dendrites seen in Fig. 1
(a) develop nearly instantaneously and apparently independently of
the field sweep rate. The images in Fig. 1 (b) and (c), where the Au
layer is below 1$\mu$m, show a cross-over behavior where the widths
of the dendrites become increasingly wider and more diffuse.

The visual MOI results are quite consistent with the magnetization
results. Shown in Fig. 3 are the virgin branches of the $M-H$ loops
for a bare thin film, a film with a 0.15-$\mu m$-thick Au coating,
and a film with a 2.55-$\mu m$-thick Au coating. Even with a thin
0.15-$\mu m$-thick Au layer, the maximum magnetization is
substantially increased because the jumps in magnetization($M$) are
greatly reduced. In the case of film with the 2.55-$\mu m$-thick Au,
the magnetization had the highest value, and the jumps in $M$ almost
disappeared. Also, generally,we did not observe any magnetic flux
noise in the $M-H$ loops of some MgB$_2$ thin films that had been
contaminated with excess metallic Mg during fabrication. Notice that
the magnetization deviates among the three films only in the field
interval 50 Oe $\leq H \leq$ 2000 Oe. This is consistent with the
previous observation that both a lower and upper threshold field
exist for the formation of flux dendrites\cite{Johansen2,Wen}.
Hence, both at very low fields and for $H$ $>$ 2 kOe, the Au coating
has little or no effect.

Figure 4 (a)$-$(d) show MO images in the remanent state of MgB$_2$
thin films at $T =$ 3.8 K with four different gold thicknesses after
first applying a large field that gave a white background in the
images up to the point where maximum penetration occurred. In Fig. 4
(a) one sees a typical image of a bare MgB$_2$ thin film containing
many dendrites of two polarities. The dendrites with a dark rim
enter the film while the field is decreasing, and contain flux of
opposite polarity. The cores of these dendrites look gray and
correspond to moderate densities of antiflux. Thus, the dark rim
represents an annihilation zone where the flux density is exactly
zero. In Fig. 4 (b) and (c), the number of dendrites is smaller than
in Fig. 4 (a), while in Fig. 4 (d) the dendrites are completely
absent. This means that a gold layer suppresses the dendritic
instability for decreasing applied field.

Suppression of dendritic avalanches by a gold layer, and the clear
sensitivity of dendritic avalanches to the thickness of the gold
layer suggests a thermal origin for the instability. Indeed, the
gold layer increases the effective conductivity of the MgB$_2$ $+$
Au sandwich by a value proportional to the gold thickness. Higher
conductivity is known to improve the stability of a superconductor
with respect to thermal avalanches\cite{Aranson2,Rakhmanov}. The
gold conductivity at 4 K is comparable to the flux-flow conductivity
of MgB$_2$. Hence, a noticeable decline in the number of avalanches
is expected when the gold layer becomes approximately as thick as
the MgB$_2$ film. This is exactly what we found experimentally.

In summary, magnetization measurements, as well as visualizations of
flux distributions by using MOI, demonstrate that the dendritic flux
instability also exists in ultra clean carbon-free $\it c$-axis
oriented MgB$_2$ films. Our results show that the instability cannot
be caused by electron scattering due to carbon doping as argued in
Ref. 11, and fully support for a thermal origin for dendritic
avalanches. Deposition of gold on top of the MgB$_2$ film completely
suppresses the instability if the gold layer is thicker than 2.5
$\mu$m. The present work further proves that coating with a thin
metallic layer can enhance the critical current density of MgB$_2$
films at $T$ $<$ 15 K, and for $H$ $<$ 1 kOe.

\begin{acknowledgments}
This work is supported by the Creative Research Initiatives of the
Korean Ministry of Science and Technology, and FUNMAT@UiO and the
Research Council of Norway.
\end{acknowledgments}

\begin{figure}
\caption{Zero-field-cooled magnetization at 4 Oe versus temperature
(5 $\leq$ $T$ $\leq$ 42 K) of a bare MgB$_2$ thin film. The inset is
a wavelength dispersive spectrum of an arbitrary surface of a
MgB$_2$ thin film and was obtained using EPMS}
\end{figure}

\begin{figure}
\caption{Magneto optical (MO) images of flux penetrations into the
virgin states of MgB$_2$ thin films at 3.8 K for gold thicknesses of
(a) 0, (b) 0.2, (c) 0.9, and (d) 2.55 $\mu m$. The images were taken
at an applied field of 34 mT.}
\end{figure}

\begin{figure}
\caption{Initial magnetization hysteresis ($M-H$) loop for 0-,
0.15-, and 2.55-$\mu m$-thick Au films on MgB$_2$ thin films at 5 K
in the field range of 0 $\leq$ $H$ $\leq$ 2000 Oe.}
\end{figure}

\begin{figure}
\caption{Remanent state after a maximum applied field of 60 mT for
MgB$_2$ thin films at 3.8 K with gold thicknesses of (a) 0, (b) 0.2,
(c) 0.9, and (d) 2.55 $\mu m$.}
\end{figure}

\end{document}